\documentclass[namedreferences]{solarphysics}

\usepackage[optionalrh,solaromanenum,natbib]{spr-sola-addons} 
\usepackage[dvips]{graphicx}                    
\usepackage{amssymb}                    
\usepackage{color}                       
\usepackage{url}                         
\usepackage[utf8]{inputenc}
\usepackage{multirow}

\definecolor{grey}{rgb}{0.5,0.5,0.5}
\newcommand{\etal}{{\it et al.}}

\begin{document}
\begin{article}
\begin{opening}

\title{Revisiting the frequency drift rates of decameter type III solar bursts observed in July--August 2002}
\author{A.A.~\surname{Stanislavsky}$^{1}$\sep
        A.A.~\surname{Konovalenko}$^{1}$\sep
        E.P.~\surname{Abranin}$^{1}$\sep
        V.V.~\surname{Dorovskyy}$^{1}$\sep
        A.~\surname{Lecacheux}$^{2}$\sep
        H.O.~\surname{Rucker}$^{3}$\sep
        Ph.~\surname{Zarka}$^{2}$
       }

\runningauthor{Stanislavsky \textit{et~al.}} \runningtitle{Revisiting the frequency drift rates of decameter type III solar bursts}

  \institute{$^1$ Institute of Radio Astronomy, National Academy of Sciences of Ukraine, Kharkiv, Ukraine\\
                     email: \url{a.a.stanislavsky@rian.kharkov.ua}\\
             $^{2}$ LESIA \& USN, Observatoire de Paris, CNRS, PSL/SU/UPMC/UPD/SPC/UO/OSUC, Meudon, France\\
             $^{3}$ Commission for Astronomy, Austrian Academy of Sciences, Graz, Austria\\
             }

\begin{abstract}
Estimating for the frequency drift rates of type III solar bursts is crucial for characterizing their source development in solar corona. According to Melnik \textit{et al.} (\textit{Solar Phys.} \textbf{269}, 335, 2011), the analysis of powerful decameter type III solar bursts, observed in July--August 2002, found a linear approximation for the drift rate versus frequency. The conclusion contradicts to reliable results of many other well-known solar observations. In this paper we report on the reanalysis of the solar data, using a more advanced method. Our study has shown that decameter type III solar bursts of July--August 2002, as standard type III ones, follow a power law in frequency drift rates. We explain possible reasons for this discrepancy.
\end{abstract}

\end{opening}

\section{Introduction}
Solar type III bursts have been observed since the fifties of the last century \citep{wild50}. This is the largest population of bursts, and it would seem that they are the most studied events from observational and theoretical points of view \citep{kruger79}. Nevertheless, interest in the study of such bursts does not vanish, but even rather \textit{vice versa} \citep[and references therein]{rr14}. The manifestation of solar activity is detected in a very wide frequency range, from several GHz down to a few tens of kHz. They are generated by fast electron beams propagating with velocities of about 0.3$c$ ($c$ is the velocity of light) along open magnetic field lines in the solar corona. The fast electrons induce Langmuir waves along the beam propagation path, and the waves are scattered on ions and transformed into radio emission \citep[\textit{e.g.,}][]{zhel70}. The most detailed and most recent analysis of type III burst properties with their interpretation can be found in the recent review \citep{rr14}. The radiation mechanism is nonthermal, and therefore its radiation intensity increases with decreasing frequency. The flux in different solar type III bursts can vary in significant limits, which are even difficult to designate. The strong events are observed with help of both large (such as the UTR-2) and small (e-Callisto, for example) radio telescopes, whereas weak type III bursts are only detected with large instruments. The study of the phenomena does not cease to amaze. For example, several decameter type III bursts can have frequency drift rates with variable sign \citep{Melnik15}. A possible interpretation of the events is that in some frequency ranges the group velocity of radio emission is smaller than the velocity of beam electrons responsible for this radio emission. Therefore, the burst radiation at a lower frequency is received earlier than at a higher frequency. On the other hand, \cite{Stanislavsky16} have shown that the solar bursts, observed on 19 August 2012, with high-frequency cut-off are nothing else but the type III radio bursts. This group of solar bursts had frequency drift rates and durations of individual events similar to the features of type III radio bursts at low frequencies. The appropriate interpretation of solar bursts with high-frequency cut-off is that their radiating sources move behind the solar limb relative to an observer on Earth through tunnel-like cavities with low density. Further, one can expect frequency-time distortions of received type III bursts. All this requires a more careful analysis of their measurements.

An important parameter of type III solar bursts is their frequency drift rate. It characterizes the velocity of burst sources (electron beams) in the solar corona. In particular, the value is much higher, for example, than one for type II bursts that originate from shock waves. For many solar type III bursts within 75 kHz to 550 MHz \cite{ah73} obtained the equation of frequency drift rates in the form $df/dt = - (0.01 \pm  0.008)\,f^{1.83 \pm 0.39}$. Then, from observations of a group of solar type III radio bursts on December 27, 1994 by the radiospectrometer (40--800 MHz) together with the WAVES and URAP instruments \cite{mann99} have found a similar relation $df/dt=-0.0074\,f^{1.76}$. Solar type III radio observations in the range 125 kHz -- 16 025 kHz from the STEREO/\textit{Waves} experiment indicate in favor of the frequency drift rate as a function of frequency: $df/dt = - 0.02\,f^{1.80\pm 0.03}$ \citep{vm09}. The Potsdam data from 1997 to 2003 confirm again a power law of frequency drift rates in type III solar bursts \citep{Cairns09,Lobzin10}. In a very recent work \cite{rk18}, using a selection of type III bursts observed in the frequency range 30--70 MHz by LOFAR between April–September 2015, have shown that the drift rate of the bursts can also be fitted by a power-law function of frequency. On the other hand, \cite{Melnik11} note that the connection between drift rate and frequency for powerful solar type III bursts for each day of observations in July--August 2002 is linear, $df/dt=-Af+B$, where $A$ and $B$ are constants varying from day to day, and the contribution of $B$ in $df/dt$ attains 33.3 $\%$ in 10 MHz and 12.2 $\%$ in 30 MHz. It should be noted that this study was carried out at the mean frequencies 11.5, 14, 17.5, 22.5 and 27.5 MHz. In this way 163 bursts of July 2002 and 231 bursts of August 2002 have been analyzed. To cover the frequency range 10--30 MHz, the 60-channel spectrometer was used. The analog multichannel receiver was tuned to selected 60 frequencies with the frequency bandwidth 3 kHz in each frequency channel. However, the frequency channels had non-uniform spacing in frequency, and the frequency gaps between neighbor frequency channels in the spectrometer were from 110 kHz to 1.4 MHz depending on the radio interference environment. Moreover, the selected 60 frequencies changed during the observations due to sporadically occurring interferences, but these changes were not taken into account. We think that it is useful to analyze the observations again, using more advanced tools, modern numerical receivers. It should be mentioned also that the results of many years (1973--1984) of observations with the UTR-2 telescope in 12.5--25 MHz have shown that the rate of frequency drift depends little on the phase of the 11-year activity cycle and was proportional to the 1.7 power of the frequency \citep{abt90}.

In this paper we revisit the results of observations in July--August 2002, discussed previously by \cite{Melnik10} and \cite{Boiko11a,Boiko11b}. Our analysis is based on the data made with the digital spectral polarimeter, DSP \citep{Lecacheux98,Lecacheux04}. These measurement were carried out simultaneously with the 60-channel spectrometer records, but the DSP data were not used to determine the frequency drift rate of solar type III bursts before. With help of the additional measurements we are going to confirm or not the law of frequency drift rates in frequency, presented in \cite{Melnik11}.

\section{Observations and Facilities}\label{par2}
Features of the summer observational campaign of 2002 with the radio telescope UTR-2 (Kharkov, Ukraine) were described in detail by \cite{Melnik10}. Here we recall only some of them for the convenience of the reader. As for general facilities of the UTR-2, they are presented in detail in the comprehensive review by \cite{Konov16}. To observe solar radio emission, three sections of the radio telescope with a total area of 30 000 m$^2$ formed a beam with angular sizes 10$^\circ$ $\times$ 13$^\circ$, covering the solar corona at decameter wavelengths. The digital receiver DSP recorded the data in the continuous frequency band of 12 MHz (18 -- 30 MHz) with frequency resolution of 12 kHz and time resolution of 100 ms \citep{Lecacheux98,Lecacheux04}. In comparison with the 60-channel receiver the number of frequency channels of the DSP is much higher, 1024. This allows us to get noticeably more data points for the analysis of frequency drift rates and other parameters.

\begin{table}[h]
\caption{The most intensive solar flares observed in July-August 2002.}
\centering
\begin{tabular}{llp{5.2cm}c c c c c c}
\hline\hline
Date  &  Class  &  Region   &    Start    &    Max   & End \\
\hline
2002-Jul-15	&	X3.0  &	10030 &	19:59 & 20:08 & 20:14 \\
2002-Jul-23	&	X4.8  &	10039 &	00:18 & 00:35 & 00:47 \\
2002-Jul-20	&	X3.3  &	10039 &	21:04 & 21:30 & 21:54 \\
2002-Aug-24	&	X3.1  &	10069 &	00:49 & 01:12 & 01:31 \\
\hline
\end{tabular}
\label{tab1}
\end{table}

\begin{figure}
  \centering
    \includegraphics[width=1\textwidth]{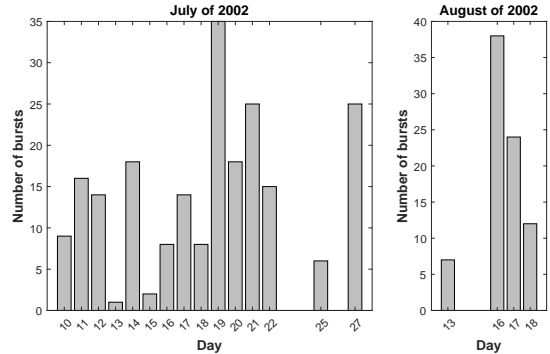}
    \caption{Distributions of powerful type III bursts in July and August 2002.}
    \label{fig1}
\end{figure}

During July--August 2002, the Sun made about two rotations. Many active regions (AR) were observed on the solar disk, from AR 10008 to AR 10096. The number of ARs varied from 4 to 13 in July, and from 7 to 14 in August. According to an overview of the strongest solar flares since June 1996 (top 50 solar flares, https://www.spaceweatherlive.com/en/solar-activity/top-50-solar-flares), four events were noticed in July--August 2002 (see Table~\ref{tab1}). The greatest manifestation of solar activity was connected with ARs 10017, 10030,   10039 and 10069. Likely, they were responsible for strong solar type III radio bursts observed at that time. The other ARs were not so effective. Using the DSP, we have analyzed 214 bursts of July 2002 and 81 bursts of August 2002. Their temporal distribution (in days) is shown in Figure~\ref{fig1}. In the case when radio time profiles of different type III bursts overlapped so that they could not be distinguished from one another, the overlapping bursts were not taken into account in this analysis, because it was difficult to track their peaks separately.

\section{Data analysis}\label{par3}
\begin{figure}
  \centering
    \includegraphics[width=1\textwidth]{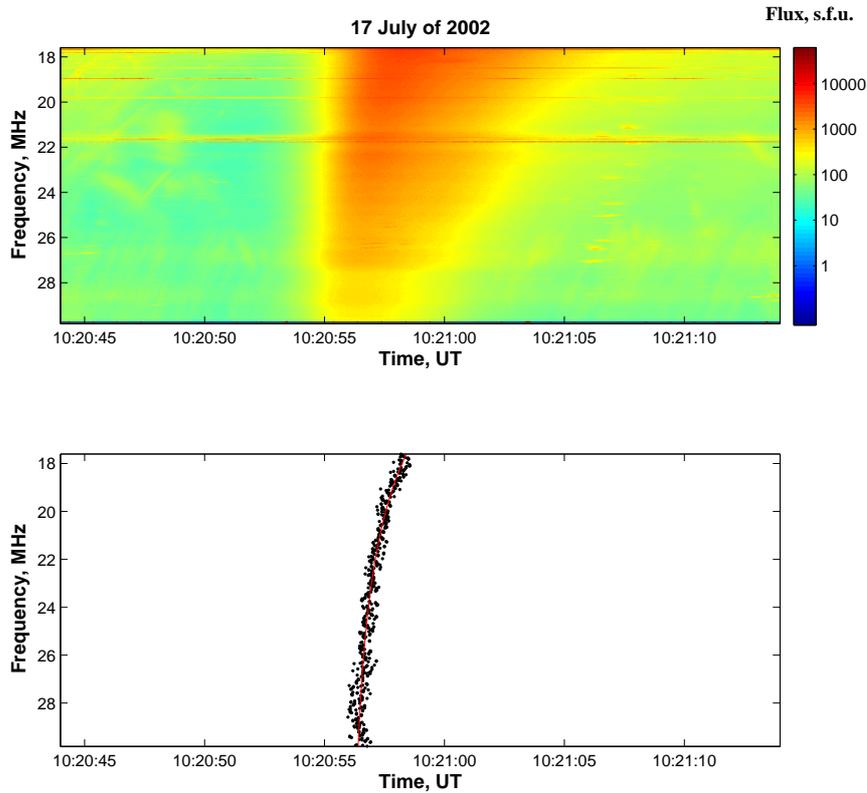}
    \caption{Example of applying the fitting procedure to time--frequency traces of type III burst peaks.}
    \label{fig2}
\end{figure}

To explore the frequency drift rates of solar type III radio bursts in July--August 2002, we examined the maximum of the power spectral density time profiles for most of the bursts and at each of the frequency channels, if possible, between 18 and 30 MHz. As usual, frequency channels of the digital records, clogged by radio narrow-band interferences, were ignored. Figure~\ref{fig2} shows the dynamic radio spectrum of such an event recorded with the UTR-2. The frequency drift rate behavior is in direct connection with the mechanisms leading to the emission of type III bursts. The type III bursts have a negative frequency drift rate, because of their sources (streams of fast electrons) originating from the Sun, traveling outward thorough the solar corona, producing radio waves at frequencies equal to the fundamental and to the second harmonic of the local plasma frequency, and the background density of solar plasma decreases as a function of distance from the Sun.

\begin{figure}
  \centering
    \includegraphics[width=1\textwidth]{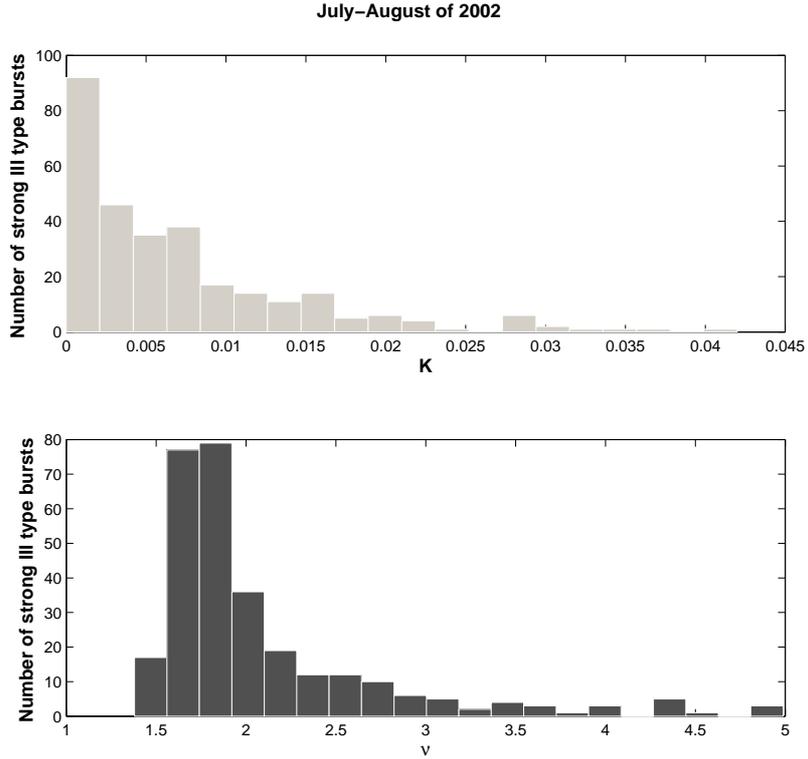}
    \caption{Histograms of the parameters $K$ and $\nu$, characterizing a power law of frequency drift rates of type III bursts.}
    \label{fig3}
\end{figure}

As a fitting function, it is convenient to take the form
\begin{equation}
f(t) = a(t-b)^{-\alpha}\,, \label{eq1}
\end{equation}
where $a$, $b$ and $\alpha$ are parameters giving the best fitting result, if one draws a line through the peak flux value of each type III burst on its dynamic spectrum in each frequency channel (see the bottom panel of Figure~\ref{fig2} as an example). The representation has a clear astrophysical interpretation. If the emission source moves in the solar corona with constant velocity $v_b$ and negligible acceleration, it travels a distance equal to $r(t) =r_0+v_b(t-t_0)$, where $r_0$ is the initial position, $t_0$ being the starting time \citep{Cairns09,Lobzin10}. Generally, the beam velocity is not constant throughout the entire propagation path of electron beams through the solar corona and interplanetary space from the Sun. Nevertheless, the assumption that any change in $v_b$ was small at the heights, where the radio emission of solar type III bursts in the frequency range 10--30 MHz was generated, is justified and can therefore be ignored. The background electron density $n_e(r)$ at the source location can be characterized by the local plasma frequency $f_p(r) = C(r/R_s-1)^{-\alpha}$, where $R_s$ is the solar radius. Then, the source produces radio emission with the frequency drift in time, $f(t) = mf_p(r(t))$, according to Equation \ref{eq1} with $a = mC(R_s/v_b)^\alpha$ and $b = t_0 + (R_s - r_0)/v_b$, where $m=1$ and $m=2$ for the fundamental and harmonic radio emissions of the given burst, respectively. Consequently, we obtain Equation \ref{eq1} in which constants $a$ and $b$ are dependent on constants $C$, $r_0$, $t_0$, $v_b$ and $m$. Next, let us write the frequency drift rate as
\begin{equation}
\frac{df}{dt} = - K\,f^\nu\,,\label{eq2}
\end{equation}
where $K$ and $\nu$ are constants. Note that the values are found directly from the recorded spectra of type III bursts. By the differentiation of Equation \ref{eq1} and the change of variables, we derive the relationship between the constants of Equations \ref{eq1}--\ref{eq2}, viz. $K=\alpha a^{-1/\alpha}$ and $\nu=1+1/\alpha$. 
Based on the processing procedure, we have analyzed frequency drift rates of the solar type III bursts. The fitting errors of $K$ and $\nu$ of each burst individually were not taken into account, as they were considered negligible. Our data statistics is shown in the histograms of Figure~\ref{fig3}. According to the histograms, the values $K$ and $\nu$ of the data set have skewed distributions. They are characterized by the mean, mode and median (Table~\ref{tab2}). To avoid the standard deviation being larger than the mean, one can consider $K$ as $10^{-x}$, in a similar way to \cite{ah73}. In this case we have $x = 2.67 \pm 1.11$. It should be noticed that the power dependence (Equation \ref{eq2}) of frequency drift rate in frequency is not only characteristic of solar type III bursts, but also of solar drift pairs \citep{Stanislavsky17}, S bursts \citep{McConnell82} and even type II bursts \citep{agmykt05}.

\begin{table}[ht]
\caption[]{Statistical properties of the solar type III bursts in frequency drift rates for parameters shown in the histograms in Figure~\ref{fig3}.}
\begin{tabular}{|*{8}{c|}}\hline
\multicolumn{4}{|c|}{$K$} & \multicolumn{4}{|c|}{$\nu$}\\ \hline
mean & mode & median & std & mean & median & mode & std\\ \hline
0.0069 & 0.0051 & 0.0032 & 0.0075 & 2.1 & 1.85 & 1.66 & 0.66 \\ \hline
\end{tabular}
\label{tab2}
\end{table}

\begin{figure}
  \centering
    \includegraphics[width=1\textwidth]{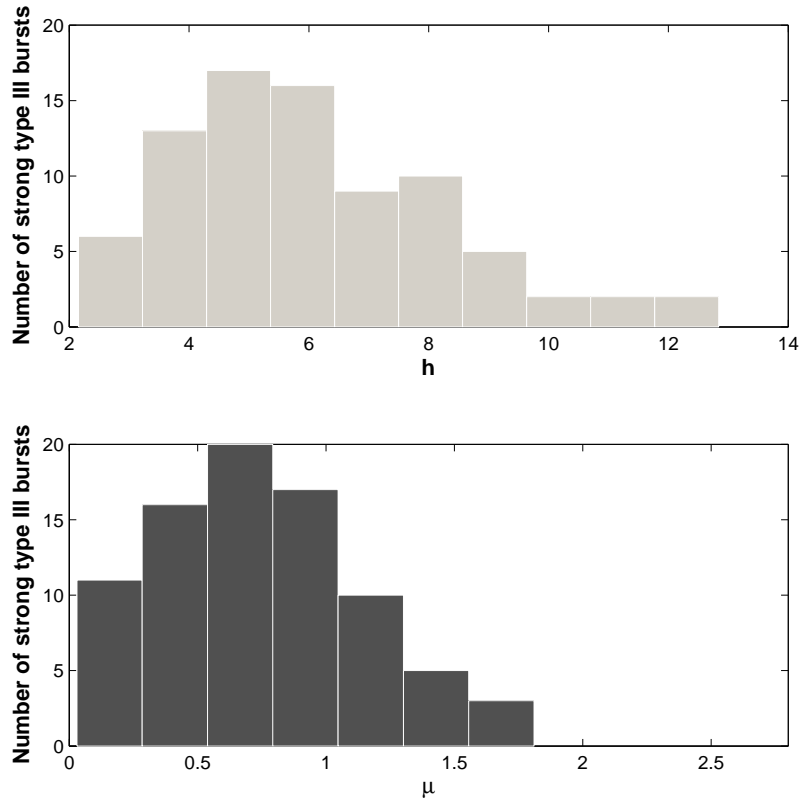}
    \caption{Histograms of the parameters $h$ and $\mu$, describing a power law of durations of type III bursts.}
    \label{fig4}
\end{figure}

\begin{table}[ht]
\caption[]{Statistical properties of the solar type III bursts for parameters shown in the histograms in Figure~\ref{fig4}.}
\begin{tabular}{|*{8}{c|}}\hline
\multicolumn{4}{|c|}{$h$} & \multicolumn{4}{|c|}{$\mu$}\\ \hline
mean & mode & median & std & mean & median & mode & std\\ \hline
6.07 & 2.15 & 5.76 & 2.26 & 0.743 & 0.7 &  0.03 & 0.42\\ \hline
\end{tabular}
\label{tab3}
\end{table}

Commonly, the duration of type III bursts, using the full-width half-maxi-mum, increases as frequency decreases \citep{wild50}. \cite{elgaroy72} have analyzed the duration of type III bursts in a wide frequency range, from 300 kHz to 500 MHz, obtained by different radio instruments. The best fit between their observed frequency $f$ in MHz and duration $\tau$ is described by the following relationship
\begin{equation}
\tau = h\,\left(\frac{f}{30\,{\rm MHz}}\right)^{-\mu}\label{eq3}
\end{equation}
with $h=6.2$, $\mu = 2/3$, and \cite{rk18} reported $\tau = (3.7 \pm 0.2)/f^{0.86\pm 0.11}$ (per 30 MHz) for 31 selected events. As applied to our data set, the dependence $\tau(f)$ is fitted as $\tau = (6.07 \pm 2.26)/f^{0.743 \pm 0.42}$ (per 30 MHz). Fitting errors of individual bursts were small and ignored. When the type III bursts had a tendency to occur in a group, we ignored such events because of their overlapping of radio time profiles. Distributions of $h$ and $\mu$ for the data set are shown in Figure~\ref{fig4}. They are skewed and have a long tail. Their statistical features are presented in Table~\ref{tab3}. Returning to the paper of \cite{Melnik11}, we can find a significantly different conclusion: at high frequencies (near 30 MHz) the 60-channel observational data are closer to the approximation $\tau=200/f$ taken from \cite{wild50}. However, this result is not confirmed by the DSP data. Our analysis of frequency drift rates and durations in the solar type III bursts, recorded by the DSP in July--August 2002, shows that the solar bursts has clear signatures typical for well-known standard type III bursts \citep[etc]{ah73,Lobzin10,rr14}.

\section{Conclusions}\label{par5}
In this paper we have shown that the analysis of solar type III radio bursts, observed in July--August 2002, with help of the DSP data does not confirm a linear character of their frequency drift rates in dependence of frequency that was reported in \cite{Melnik11}. Withal, our results are in very good agreement with \cite{ah73} who gave an empirical expression for the frequency drift rate as a power function of frequency. It is important because the dependence seems to be valid for both weak and strong bursts. This stresses that in both cases the electron beam speed dependence on distance from the Sun is weak, and it can be extrapolated from model representations. The solar corona is a dynamic structure. The background density gradient is almost unchanged on timescales comparable to the electron beam velocity but will be different from burst to burst. It is well known that active regions are directly responsible for type III radio bursts. Their drift rate increases during the approach of the active region to the central meridian. Therefore, the standard deviations for the exponents $\nu$ and $\mu$ can be quite large. Note also that some values of $\nu$ are exceptionally large, leading to large frequency drift rates. Having a smaller duration, the bursts occurred most often when the active region AR 10030 was near the central meridian. Following \cite{Ledenev00}, if the electron velocity is close to the group velocity of the electromagnetic waves, then the drift rate may be very large. When fluxes of the solar type III bursts were more 1000 s.f.u. (1 s.f.u. = 10$^{-22}$ W m$^{-2}$ Hz$^{-1}$) at low frequencies of the observations and became less at higher frequencies, we did not find any noticeable change in the frequency drift rate law. A possible explanation of discrepancy between our conclusions and ones of \cite{Melnik11} is that in the latter case the data set for solar type III radio bursts were obtained basically by the analog multichannel receiver with the frequency bandwidth 3 kHz in each frequency channel. Consequently, the data contained notable omissions. In addition, the whole frequency band from 10 to 30 MHz was divided into some frequency sub-bands 10--13, 13--15, 15--20, 20--25 and 25--30 MHz, \textit{i.\,e.} the statistical analysis was conducted on the average frequencies of these sub-bands. A simple straight line is not complex enough to accurately capture relationships between the frequency drift rates of solar type III radio bursts and their frequency. In fact, such a model is underfitted. One of the most effective ways to avoid underfitting is making sure that a model is sufficiently complex and the most appropriate for corresponding data set. The same applies to the duration of the solar type III bursts in July--August 2002.

\begin{acks}
This research was partly supported by Research Grant 0118U000563 from the National Academy of Sciences of Ukraine. We thank the anonymous referee for useful comments.
\end{acks}
\\

\noindent\textbf{Disclosure of Potential Conflicts of Interest} The authors declare that they have no conflicts of interest.

\end{article}
\end{document}